\documentclass[fleqn,twoside]{article}
\usepackage{espcrc2}

\newcommand{\cL}{{\cal L}}
\newcommand{\qbar}{{\overline{q}}}
\newcommand{\qt}{{\tilde q}}
\newcommand{\cV}{{\cal V}}
\newcommand{\cG}{{\cal G}}
\newcommand{\Psibar}{{\overline{\Psi}}}
\newcommand{\chibar}{{\overline{\chi}}}
\newcommand{\phibar}{{\hat{\phi}}}
\newcommand{\str}{{\rm str\,}}

\newcommand{\AmS}{{\protect\the\textfont2
  A\kern-.1667em\lower.5ex\hbox{M}\kern-.125emS}}

\title{On the phase diagram of quenched QCD with Wilson fermions}

\author{Maarten Golterman\address{Department of Physics and Astronomy,\\
        San Francisco State University, \\ 
        1600 Holloway Ave, San Francisco, CA 94044, USA}%
        \thanks{e-mail: {\tt maarten@stars.sfsu.edu}}}
       
\begin{document}

\begin{abstract}
In this talk, I reported on recent work on the Aoki phase diagram
for quenched QCD with two flavors of Wilson fermions.  Part of
this work was done in collaboration with Yigal Shamir \cite{gspaper},
and a shorter account of this part appeared recently \cite{gslat2003}.
In this write-up, I will therefore limit myself to reporting
on work done with Steve Sharpe and Robert Singleton, Jr.
\cite{gss}, which has not been published yet.  
We discuss the symmetries of quenched QCD,
paying careful attention to non-perturbative issues.  This
allows us to derive an effective lagrangian which agrees with
standard quenched chiral perturbation theory, but which can
also be used to address questions of a non-perturbative nature.
\vspace{1pc}
\end{abstract}

\maketitle

\section{Introduction}

The original aim of the work I report on in this talk was to see
whether a phase-diagram
analysis based on the effective theory describing the
Goldstone-boson physics of full QCD with two flavors of Wilson
fermions \cite{shsi} can be extended to the quenched case as well.
The unquenched theory is expected to have a so-called Aoki
phase \cite{aoki} at non-zero lattice spacing, and Ref.~\cite{shsi}
demonstrated that this can be understood from an effective-%
lagrangian point of view, if one includes the effects of
scaling violations in a systematic, Symanzik-like expansion
in powers of the lattice spacing (see also Ref.~\cite{creutz}).

This investigation led to two different problems which needed
to be resolved in order to study the possible
existence of an Aoki phase
in quenched QCD.  First, it was observed numerically (see
for example Ref.~\cite{scri}) that there always appears to
exist a non-zero density of near-zero modes of the (hermitian)
Wilson--Dirac operator in quenched QCD if the quark mass is
in the super-critical region.\footnote[1]{Defined as the region
where the Wilson--Dirac operator {\it in principle} can have
exact zero modes.}  Through the Banks--Casher relation \cite{bc}
this would seem to lead to the conclusion that an Aoki 
condensate always exists in the supercritical region,
in contrast to what is expected in unquenched QCD.  It was
shown in Ref.~\cite{gspaper} that this is not the case,
if one defines the Aoki phase as that region of the phase
diagram where Goldstone bosons associated with the symmetry
breaking occur.  It was argued that regions may exist where
the condensate does not vanish because of the existence of
a density of {\it exponentially localized} near-zero modes, {\it without}
any of the corresponding long-range physics usually associated
with spontaneous symmetry breaking (SSB).  This phenomenon
is an artifact of the quenched approximation.  It follows
that an effective-theory investigation of the phase structure
following Ref.~\cite{shsi} should also be possible for the
quenched theory, and it is that aspect of the project that I
will report on here.

However, in setting up an effective lagrangian for the
quenched theory which is also suitable for non-perturbative
questions such as the phase structure of the theory, one
runs into the second problem.  The leading-order
effective lagrangian in standard quenched chiral perturbation
theory \cite{bgq} contains a potential (the term linear
in the quark-mass matrix) which has a saddle point extremum
at $\Sigma=1$, where $\Sigma$ is the $U(N|N)$ valued
non-linear field describing the Goldstone multiplet of
quenched QCD (with $N$ flavors).  However, this saddle-point
is not a minimum of the potential.  Worse, if one takes
this effective lagrangian literally, one would conclude
that the condensate in the ghost-quark sector is different
from that in the physical quark sector.  Clearly, this is
nonsense, since the ghost-quark sector is only introduced
to give a path-integral definition of quenched QCD --
ghost quarks have no business of living a life of their own.

It turns out that a careful reanalysis of the symmetries
of quenched QCD (in particular those of the ghost sector)
\cite{damgaardetal} leads to a resolution of the second problem.
While chiral perturbation theory (ChPT) starting from
the effective lagrangian of Ref.~\cite{bgq} (and
expanding around the saddle point $\Sigma=1$) is the
same as that derived from the effective lagrangian
constructed here (and in Ref.~\cite{gss}) \cite{shsh}, these two
effective lagrangians are different.  The one I will
discuss here makes it possible to investigate the 
existence of an Aoki phase in quenched QCD following
the lines along which this was done for unquenched
QCD \cite{shsi}.

\section{Rigorous definition of quenched QCD}

We will begin with a discussion of the lagrangian for
quenched QCD in the continuum.
The euclidean lagrangian for quenched QCD is \cite{morel,bgq}
\begin{equation}
\label{ql}
\cL=\qbar Dq+\qbar Mq+\qt^\dagger D\qt+\qt^\dagger M\qt\,.
\end{equation}
Here the Grassmann variables $q$ and $\qbar$ describe the
physical (valence) quarks, and the complex $c$-number
field $\qt$ describes the ghost quarks.  Note that,
unlike $\qbar$, $\qt^\dagger$ is not an independent
variable.  The path integral over $\qt$ and $\qt^\dagger$
is only well defined if $\qt^\dagger$ is the hermitian
conjugate of $\qt$, and if the eigenvalues of the quark-mass 
matrix $M$ have a positive real part (in euclidean space, $D$,
the Dirac operator, has purely imaginary eigenvalues).
It follows that the kinetic term of the ghost quarks
couples $\qt_L=\frac{1}{2}(1+\gamma_5)\qt$ to
$\qt^\dagger_R=\qt^\dagger\frac{1}{2}(1-\gamma_5)$.
This is unlike the physical sector, where the left- and 
right-handed projections of $q$ and $\qbar$ maybe defined
independently.

Having a convergent integral defining the ghost sector of
the theory is a step closer to defining the full path
integral for quenched QCD non-perturbatively.  However,
the path integral still needs to be regularized, and
we will do so here employing Wilson fermions.  This is
done by replacing $D$ by the naive nearest-neighbor covariant
Dirac operator on the lattice, and by adding the Wilson mass
term
\begin{equation}
\label{wm}
\qbar Wq+\qt^\dagger W\qt\,,
\end{equation}
with
\begin{eqnarray}
\label{wmdef}
(Wq)(x)&=&-\frac{r}{2}\sum_\mu\left(
U_\mu(x)q(x+\mu)\right.\\
&&\left.+U^\dagger_\mu(x-\mu)q(x-\mu)-2q(x)
\right)\,.\nonumber
\end{eqnarray}
Now a new problem arises: the eigenvalues of $D+W+M$
may have negative real parts for certain gauge fields
if the quark masses in $M$ are close to their critical
values.  This would make the ghost-quark path integral
ill-defined.  We may fix this problem as follows.
We start over with {\it unquenched} lattice QCD with Wilson
fermions.  The path integral for this theory exists,
irrespective of the properties of the Wilson--Dirac
operator.  In this theory, we rotate each quark field:
\begin{equation}
\label{rot}
q\to e^{i\frac{\pi}{4}\gamma_5}\;q\,,\ \ \ 
\qbar\to\qbar\;e^{i\frac{\pi}{4}\gamma_5}\,.
\end{equation}
This leads to a quark lagrangian
\begin{equation}
\label{vl}
\cL_{\rm quark}=
\qbar(D+i\gamma_5(W+M))q\,.
\end{equation}
Note that the operator $D+i\gamma_5(W+M)$ is anti-hermitian,
and also that the axial rotation of Eq.~(\ref{rot}) is {\it not}
anomalous (since it rotates both $W$ and $M$).  We may now
define quenched QCD with Wilson fermions by adding for
the ghost sector the lagrangian
\begin{equation}
\label{gl}
\cL_{\rm ghost}=\qt^\dagger(D+i\gamma_5(W+M))\qt
+\epsilon\qt^\dagger\qt\,.
\end{equation}
The lagrangian $\cL_{\rm glue}+\cL_{\rm quark}+\cL_{\rm ghost}$
defines quenched QCD with Wilson fermions in the limit
$\epsilon\searrow 0$.

\section{Symmetries of quenched QCD}

In order to construct an effective theory, we need to
identify the symmetries of the theory first.  Because of
the fact that $\qt^\dagger$ is not independent of $\qt$,
it turns out that the symmetries of quenched QCD are
in fact different \cite{damgaardetal,gss,shsh} 
from those discussed in Ref.~\cite{bgq}.

It is instructive to first consider the physical and ghost
sectors separately.  Ignoring the Wilson
mass term $W$, the symmetries of the physical sector
are
\begin{eqnarray}
\label{vs}
q_{L,R}&\to& V_{L,R}q_{L,R}\,,\\
\qbar_{L,R}&\to&\qbar_{L,R} V_{L,R}^{-1}\,,\nonumber\\
V_{L,R}&\in&GL(N)\,.\nonumber
\end{eqnarray}
The symmetry group is $GL(N)_L\times GL(N)_R$ (for $N$
flavors of physical quarks).

As mentioned before, the kinetic term of the ghost quarks
couples left- and right-handed ghost quarks, and thus
the transformation of the right-handed fields determines
that of the left-handed fields:
\begin{eqnarray}
\label{gs}
\qt_R&\to&V\qt_R\ \ \ \Rightarrow\ \ \ \qt_L\to V^{\dagger-1}\qt_L\,,\\
V&\in&GL(N)\,.\nonumber
\end{eqnarray}
In other words, the symmetry group in the ghost sector is
just the group $GL(N)$.  The vector subgroup in the physical
sector (defined as the symmetries which leave the Wilson
mass term invariant) is the diagonal subgroup $GL(N)\subset
GL(N)_L\times GL(N)_R$; in the ghost sector the Wilson
term $\qt^\dagger W\qt=\qt_L^\dagger W\qt_L+\qt_R^\dagger W\qt_R$
is invariant when $VV^\dagger=1$, and the vector subgroup
is the group $U(N)\subset GL(N)$.
One may introduce spurion fields for the mass terms in order
to trace the quark-mass dependence of the effective theory
in the usual way \cite{gss}.  

Taking into account also transformations between physical
and ghost quarks, the full symmetry group is \cite{shsh,gss}
\begin{eqnarray}
\label{cs}
\cG&=&\Bigl\{(\cV_L,\cV_R)\in GL(N|N)_L\times GL(N|N)_R\nonumber\\
&&\ \ \ \ \ 
|\cV_{Lgg}|_{\rm body}=\cV_{Rgg}^{\dagger-1}|_{\rm body}\Bigr\}
\,,\\
\cV_L&=&\pmatrix{\cV_{Lqq}&\cV_{Lqg}\cr\cV_{Lgq}&\cV_{Lgg}}\,.
\nonumber
\end{eqnarray}
Here $qq$, $qg$, {\it etc.} denote the quark-quark, quark-ghost,
{\it etc.} $N\times N$ blocks of the $2N\times 2N$ matrix
$\cV_L$, and the subscript ``body" refers to the $c$-number
parts of the matrix elements of the $gg$ block.

\section{Effective theory for quenched QCD}

We start with the construction of the effective theory
in the continuum limit, using the symmetries of quenched QCD
without $W$.  As usual, we introduce a non-linear field
$\Sigma={\rm exp}(\Phi)$ transforming like $\Psi_L\Psibar_R$
(with $\Psi_L=\pmatrix{q_L\cr\qt_L}$, 
$\Psibar_R=\pmatrix{\qbar_R&\qt^\dagger_L}$, {\it etc.}):
\begin{equation}
\label{st}
\Sigma\to\cV_L\Sigma\cV_R^{-1}
\Rightarrow\Sigma^{-1}\to\cV_R\Sigma^{-1}\cV^{-1}_L\,.
\end{equation}
The field $\Phi$ can be written as
\begin{equation}
\label{phi}
\Phi=\pmatrix{i\phi_1+\phi_2&\chibar\cr\chi&\phibar}\,,
\end{equation}
where $i\phi_1+\phi_2$ parameterizes the coset
$GL(N)_L\times GL(N)_R/GL(N)$, and $\phibar$
parameterizes the coset $GL(N)/U(N)$.
The order $p^2$ invariant chiral lagrangian is then
given by \cite{gss}
\begin{eqnarray}
\label{leff}
\cL_{\rm eff}&\!\!\!=\!\!\!&\frac{1}{8}f^2
\str(\partial_\mu\Sigma\partial_\mu\Sigma^{-1})
-\!v\;\str M(\Sigma+\Sigma^{-1}).\nonumber\\
\end{eqnarray}
In order to arrive at this result, one also makes use
of a parity symmetry \cite{gss}.  

Note that the lagrangian depends only on the field
$\phi\equiv\phi_1-i\phi_2$, and not on its hermitian conjugate.
There is therefore no redundancy in the fields describing
the Goldstone sector, but in order to set up a path
integral for the effective theory, we need to specify
the contour along which this field is integrated.
We specify the contour by taking $\phi$ to be real,
{\it i.e.} we set $\phi_2=0$.  While we do not have 
a ``derivation" of this choice, it is a sensible choice
for the following reasons.  First, the same issue arises
for the euclidean path integral in the unquenched case.
There, after continuation to Minkowski space, one
imposes the condition $\qbar=q^\dagger\gamma_0$, which
restricts all $GL(N)$ groups to their unitary subgroups,
and $\Sigma$ then describes the coset $U(N)_L\times U(N)_R/
U(N)$.  Our choice in the quenched case is consistent with
this.\footnote{Most likely, it makes little sense to try
continue the quenched theory to Minkowski space.}
One can in fact demonstrate that the quenched group
integral relevant in the so-called $\epsilon$-regime 
(see for example Ref.~\cite{damgaard})
is independent of the quark mass only for this choice of
the contour \cite{jv,gss}, as the quenched partition
function should be.  Finally, we note that, expanding
around $\Sigma=1$, ChPT does not depend on the choice of
contour, and in fact coincides with standard quenched
ChPT as developed in Ref.~\cite{bgq}.

The restriction to $\phi$ real reduces the symmetry group
of the effective theory, because this condition reduces the
field space available.  However, one can show that this does
not affect the chiral lagrangian given in Eq.~(\ref{leff})
above.  For a discussion of the anomaly (which follows
closely that of Ref.~\cite{bgq}), see Ref.~\cite{gss}.

Next, we wish to extend the effective theory to incorporate
lattice spacing effects, in a systematic expansion in
powers of $a$, the lattice spacing.  I will sketch the
argument only briefly, as it follows a similar construction
for the unquenched theory very closely \cite{shsi}.
One starts with the Symanzik effective theory \cite{symanzik}
at the quark level. For the quenched theory, the relevant
quark-level effective lagrangian is
\begin{equation}
\label{lqeff}
\cL=\Psibar(D+i\gamma_5 m)\Psi+b_1\;
a\Psibar\;i\gamma_5\;i\sigma_{\mu\nu}F_{\mu\nu}\Psi+\dots\,,
\end{equation}
where $m$ is the subtracted quark mass, and $b_1$ is
the first of the Symanzik coefficients.  Note the
unusual appearance of extra factors $i\gamma_5$: they
arise because of the quark field rotation (\ref{rot})
used in the definition of the quenched theory.
The key observation is that the Pauli term which occurs
at order $a$ breaks chiral symmetry just like the mass term.
To order $m^2$, $ma$ and $a^2$, this leads to an effective-theory
potential of the form \cite{gss} (for degenerate quark masses)
\begin{eqnarray}
\label{veff}
V_{\rm eff}&=&-ic_1\str(\Sigma-\Sigma^{-1})\\
&&+c_2\left((\str\Sigma)^2+(\str\Sigma^{-1})^2\right)\nonumber\\
&&+c_3\str\Sigma\str\Sigma^{-1}\nonumber\\
&&+c_4\left(\str\Sigma^2
+\str\Sigma^{-2}\right)\,,\nonumber
\end{eqnarray}
where
\begin{eqnarray}
\label{constants}
c_1&\sim& m\Lambda^3+a\Lambda^5\,,\\
c_{2,3,4}&\sim& m^2\Lambda^2+ma\Lambda^4+a^2\Lambda^6\,,\nonumber
\end{eqnarray}
with $\Lambda$ of order the QCD scale.  The minus sign in the first
term is again a consequence of the rotation (\ref{rot}),
which changes the properties of parity symmetry \cite{gss}.
I am leaving out terms which appear as a consequence of the
anomaly, as they will not change the results discussed in the
following.

\section{Analysis of the effective potential}

While we will be interested in the two-flavor theory, let us
look at the one-flavor theory as a warm up.  
Substituting
\begin{equation}
\label{svac}
\Sigma=\pmatrix{e^{i\phi}&0\cr 0&e^\phibar}\,,
\end{equation}
the effective potential becomes, including the $\epsilon$
term of Eq.~(\ref{gl})
\begin{equation}
\label{vof}
V_{\rm eff}=2c_1\sin\phi+2ic_1\sinh\phibar+2\epsilon
\cosh\phibar+\dots\,,
\end{equation}
where the dots refer to the higher order terms in Eq.~(\ref{veff}).
We see that the effective potential is complex!
It looks like the usual procedure of minimizing a potential
to find the vacuum structure breaks down.  However, in the
context of euclidean field theory, ``finding the vacuum"
is equivalent to performing a saddle-point approximation,
and this is what we propose to do in this case as well.
For the leading-order effective potential shown in Eq.~(\ref{vof}),
the saddle-point solution is \cite{gss}
\begin{eqnarray}
\label{spsol}
\phi&=&-{\rm sign}(c_1)\frac{\pi}{2}\,,\\
\phibar&=&-i\;{\rm sign}(c_1)\frac{\pi}{2}\,.\nonumber
\end{eqnarray}
It is straightforward to check that shifting the fields
$\phi$ and $\phibar$ by this solution precisely ``undoes"
the axial rotation (\ref{rot}).  Carrying out this shift,
the effective potential becomes (for $\epsilon\to 0$)
\begin{equation}
\label{vas}
V_{\rm eff}=2|c_1|(-\cos\phi+\cosh\phibar)\,.
\end{equation}
(which has a minimum at $\phi=\phibar=0$).  Note that
the solution (\ref{spsol}), translated in terms of
quark condensates, means that
\begin{equation}
\label{qc}
\langle{\rm tr\ }q\qbar\rangle=
\langle{\rm tr\ }\qt\qt^\dagger\rangle\,,
\end{equation}
as one would expect: the ghost condensate reproduces the
physical condensate.

There is more to say about the one-flavor case, especially
if one also takes into account the anomaly.  However,
instead, I will now discuss the two-flavor case, assuming
that no non-trivial condensates form in the singlet sector.
First, we shift the singlet fields $\phi_0$ and $\phibar_0$
by the saddle-point solution (\ref{spsol}), in order to
undo the axial rotation (\ref{rot}).  We then set the
shifted singlet fields equal to zero, in accordance with
the assumption that no non-trivial singlet condensate occurs.
Then, for the two-flavor theory, we substitute
\begin{equation}
\label{svactf}
\Sigma=\pmatrix{{\rm exp}(i\sigma_3\phi/2)&0\cr 0&
{\rm exp}(\sigma_3\phibar/2)}\,,
\end{equation}
pointing the condensate in the third isospin direction
without loss of generality.  The effective potential
can now be written as
\begin{eqnarray}
\label{vefftf}
V_{\rm eff}&=&4|c_1|\left(-\cos{(\phi/2)}+\cosh{(\phibar/2)}\right)\\
&&-4(2c_2+c_3)\left(\cos{(\phi/2)}-\cosh{(\phibar/2)}\right)^2
\nonumber\\
&&+4c_4\left(-\cos\phi+\cosh\phibar\right)\,.\nonumber
\end{eqnarray}
First, consider the case $2c_2+c_3=0$.  In that case the
physical ($\phi$) and ghost ($\phibar$) sectors decouple.
In the physical sector, the potential maybe minimized,
and the solution is that found already in Ref.~\cite{shsi}
for the unquenched case.  This solution is
\begin{eqnarray}
\label{tfsol}
\phi&=&0\,,\ \ \ {\rm if}\ c_4>-|c_1|/4\,,\\
\phi&\ne&0\,,\ \ \ {\rm if}\ c_4<-|c_1|/4\,.\nonumber
\end{eqnarray}
The solution $\phi\ne 0$ corresponds to a non-zero
$\qbar\gamma_5\sigma_3 q$ condensate,\footnote{in terms
of the original quark fields before the axial rotation
(\ref{rot})} and corresponds to an Aoki phase.
From Eq.~(\ref{tfsol}), it is clear that, in order
to find a non-trivial condensate, $c_4$ and $c_1$ have
to be of the same order of magnitude.  Referring to 
Eq.~(\ref{constants}), this implies that an Aoki phase
can only occur if $am\sim a^3\Lambda^3$.\footnote{
Subtracting first the $a\Lambda^5$ term in $c_1$ ({\it cf.}
Eq.~(\ref{constants})); see Ref.~\cite{shsi} for more
detail.}  In other words,
approaching the continuum limit, the Aoki phase is
a narrow ``finger" of width $a^3$ in the $am-g$ phase diagram
\cite{shsi}.

As expected, this solution is completely reproduced in
the ghost sector, if one performs a saddle-point
extremization on the $\phibar$ part of the effective
potential.  When one turns on the coupling $2c_2+c_3$,
the solution actually remains the same, as can be 
checked from the combined saddle-point equations for
the fields $\phi$ and $\phibar$ in that case.

There is one embarrassment in the ghost sector which 
should be pointed out.  For $c_4<0$, the effective
potential is unbounded from below for large values
of the non-compact field $\phibar$.  However, one can
check that the saddle-point is still locally stable,
and we believe that therefore our physics conclusions
are valid, despite this embarrassment.  The effective
theory describes the theory only well below the
chiral-symmetry breaking scale, and clearly should
not be trusted at values of the fields above this 
scale.  It is because of this that we believe that
local stability of the saddle point solution is
sufficient to conclude that the solution in the
physical sector is reproduced in the ghost sector,
as it should.

\section{Conclusions}

In this talk, I have shown how quenched lattice QCD with
Wilson fermions can be defined rigorously.  This allows
us to decide what the symmetries of this theory are,
which then may be used to construct the effective
chiral lagrangian for quenched QCD, including lattice
spacing effects, in a systematic expansion in the quark
mass and the lattice spacing.  This precise analysis
exposes a flaw in the original construction \cite{bgq} of the
quenched effective theory, but this does not change the
perturbative expansion based on the effective lagrangian.
All ChPT results based on the lagrangian of Ref.~\cite{bgq}
therefore remain valid.  

If one wants to investigate a non-perturbative issue such
as the possible existence of an Aoki phase, the lagrangian
described here must be used.  I sketched the arguments
leading to the conclusion that the two-flavor quenched 
theory exhibits the same possible phase structure as the
unquenched two-flavor theory, which was investigated
using an effective-theory approach before \cite{shsi,creutz}.
Of course, whether the Aoki ``fingers" are realized in 
two-flavor QCD depends on the values of the coefficients
multiplying the powers of $m$ and $a$ in $c_4$, and can
thus not be answered within the framework of effective field theory.
Note also that the relevant low-energy constants could 
turn out to be different in the quenched and unquenched
cases, so that the appearance of Aoki fingers in the
quenched case does not imply their existence in the unquenched
case, and {\it vice versa}.

Finally, I skipped over many technical details,
including in particular a more thorough discussion of
the anomaly and the singlet sector.
For a more complete analysis
I refer to Ref.~\cite{gss}, which I hope will appear soon.

\section*{Acknowledgements}
I would like to thank the organizers of 
``Lattice Hadron Physics 2003" for a pleasant and 
stimulating workshop.  I also thank the Department
of Physics and the Institute for Nuclear Theory
at the University of Washington, where this work
was begun, for hospitality.
This work was supported in part by
the US Department of Energy.

\end{document}